\documentclass[]{iopart}

\usepackage{epsfig}

\begin{document}

\title{What to learn from dilepton transverse momentum spectra in heavy-ion collisions?}

\author{J\"org Ruppert}
\address{Department of Physics, 3600 Rue University, McGill University, Montreal, Quebec, Canada
H2A 2T8}
\ead{ruppert@hep.physics.mcgill.ca}

\author{Thorsten Renk}
\address{Department of Physics, PO Box 35 FIN-40014 University of Jyv\"askyl\"a, Finland, and 
Helsinki Institute of Physics, PO Box 64 FIN-00014, University of Helsinki, Finland}
\ead{trenk@phys.jyu.fi}

\pacs{25.75.-q}

\begin{abstract}

Recently the NA60 collaboration has presented high precision measurements of dimuon spectra double differential in invariant mass $M$ and transverse pair momentum $p_T$ in In-In collisions at $158~{\rm AGeV}$ \cite{NA60data}.
While the $M$-dependence is important for an understanding of in-medium changes of light vector mesons and is $p_T$ integrated insensitive to collective expansion, the $p_T$-dependence arises from an interplay between emission temperature and collective transverse flow.  
This fact can be exploited to derive constraints on the evolution model and in particular on the contributions of different phases of the evolution to dimuon radiation into a given $M$ window. We present arguments that a thermalized evolution phase with $T > 170~{\rm MeV}$ leaves its imprint on the spectra.

\end{abstract}

\section{Introduction}

The properties of vector mesons are predicted to change in the presence of a hot and dense nuclear medium. Especially interesting are the light vector mesons, namely the $\rho$, $\omega$ and $\phi$-mesons, because of their electromagnetic decay properties. Their in-medium changes are reflected in the invariant $M$-spectra of the produced dileptons and hence their spectral properties are one of the essential ingredients for the calculation of thermal dimuon emission. This emission rate needs to be convoluted with a dynamical model for the space-time evolution of matter created in nucleus-nucleus collisions. In addition, dilepton emission from a partonic phase and from four-pion annihilation are important thermal sources, in addition to non-thermal sources such as resonance decays or primordial Drell-Yan processes.

A detailed description of the dynamical evolution model employed here is given in \cite{ourModel}. The in-medium changes in the heat bath of the $\rho$ and $\omega$ at finite temperature and density are obtained by calculating the self-energies from the real and imaginary parts of the forward scattering amplitude \cite{Kapusta}. The $\phi$-meson's in-medium modifications are not yet included into this framework. Four pion-annihilation processes are calculated by Lichard \cite{Lichard} in an hadronic approach which is constrained e.g. by the recent BaBar data on lepton annihilation into four pions \cite{BaBar}.  We employ a quasi-particle model to calculate contributions from a thermalized partonic evolution phase \cite{ourModel}. This approach is successful in describing the $p_T$ integrated invariant $M$-spectrum as well as the $M$-shape for the low and high $p_T$ data selection \cite{ourModelM}. 
While the success of the model to describe the shape in invariant mass allows for the conclusion that the $\omega$- and $\rho$-vector meson are considerably broadened in the nuclear medium \footnote{Note that  
scenarios with considerable downward mass shift are disfavored by the data \cite{BRruppert}.}, further information can be gained from understanding details of the transverse $p_T$ spectra in different mass windows.
This is not only a consistency test for the model and the freeze-out prescription, but also allows for a better understanding of the dominant source in the mass region $M>1 ~{\rm GeV}$ \cite{ourModelPT}.

\section{Fireball evolution and transverse flow}

The dilepton spectrum is calculated as a convolution of the dynamical evolution with the local rate

\begin{eqnarray}
\label{E-1}
\frac{d^3N}{dM dp_T d\eta} = {\rm evolution} \otimes \frac{dN}{d^4 x d^4q} 
\end{eqnarray}
which is inferred from the averaged virtual photon spectral function $R(q, T, \rho_B)$ as 
\begin{eqnarray}
\label{E-Rate}
\frac{dN}{d^4 x d^4q}  =  \frac{\alpha^2}{12\pi^4} \frac{R(q, T, \rho_B)}{e^{p_\mu u^\mu/T}- 1},
\end{eqnarray}
 here $q$ is the four-momentum of the emitted muon pair, $T$ is the temperature 
and $\rho_B$ is the baryon density, while the fireball evolution encodes information on  the radiating volume, namely temperature, baryon density, local transverse flow and longitudinal rapidity as well as 
non-equlibirum properties such as pion- and kaon chemical potentials. Finite lepton masses are accounted for by the appropriate additional factor.
The role of transverse flow is determined by evaluating $u_\mu p^\mu$, see \cite{ourModelPT,Sollfrank}.
The temperature $T$ in Eq. (\ref{E-Rate}) together with the flow field entering via $u_\mu p^\mu$ sets the basic scale possibly modified by the momentum (and density) dependence of the spectral function $R(q,T,\rho_B)$.
In the absence of flow and for moderate momentum and density dependences of the spectral function the thermal part of the dilepton $p_T$-spectra at a given mass would occure simply as a superposition of contributions from different $T$ weighted by the radiating volume at that temperature into the acceptance. In such a scenario the low $p_T$ region would be dominated by late, low $T$ emission
wheras at larger $p_T$ the harder slopes of earlier emission from hotter regions will gradually become visible. The presence of a flow field (which is essential in any realistic description of the evolution) changes this picture.
Since initially transverse flow $\rho_T$ is small and $T$ is large whereas at late stages the situation is reversed, there are no cold emission sources without flow distortion, but hot emission regions are (almost) unaffected by flow. If $\rho_T$ is  larger at fixed $T$ the $p_T$-spectrum becomes blueshifted and hardens, at some $p_T$ (as a function of $M$ and $\rho_T$) this hardening becomes maximally and then drops again. Since electromagnetic spectra are integrated quantities over the evolution, this effect has to be quantitatively averaged. 
An additional contribution to the $p_T$-spectra can emerge from the decays of light-vector mesons in vacuum after freeze-out. For this contribution we calculate the momentum spectrum of the $\rho$-mesons after decoupling using the Cooper-Frye formula, for details see \cite{ourModel,ourModelPT}.

\section{Comparison to the NA60 Data}

The NA60 collaboration has measured $p_T$-spectra in three different mass windows, a lower mass window with $0.4~{\rm GeV}<M<0.6~{\rm GeV}$, the $\rho$-like window with $0.6~{\rm GeV}<M<0.9~{\rm GeV}$ and a higher mass window $1.0~{\rm GeV}<M<1.4~{\rm GeV}$. 
We present a calculation of both spectra and effective temperature  in Fig. 1 as inferred from the model \footnote{Measured flow spectra can locally be described by an effective temperature $T^*$ via a fit of the form $1/m_T \frac{dN}{dm_T} \simeq A\exp[-\frac{m_T}{T^*}]$ with $m_T^2=p_T^2+m^2$.}. 

\begin{figure}[htb]
\begin{center}
\epsfig{file=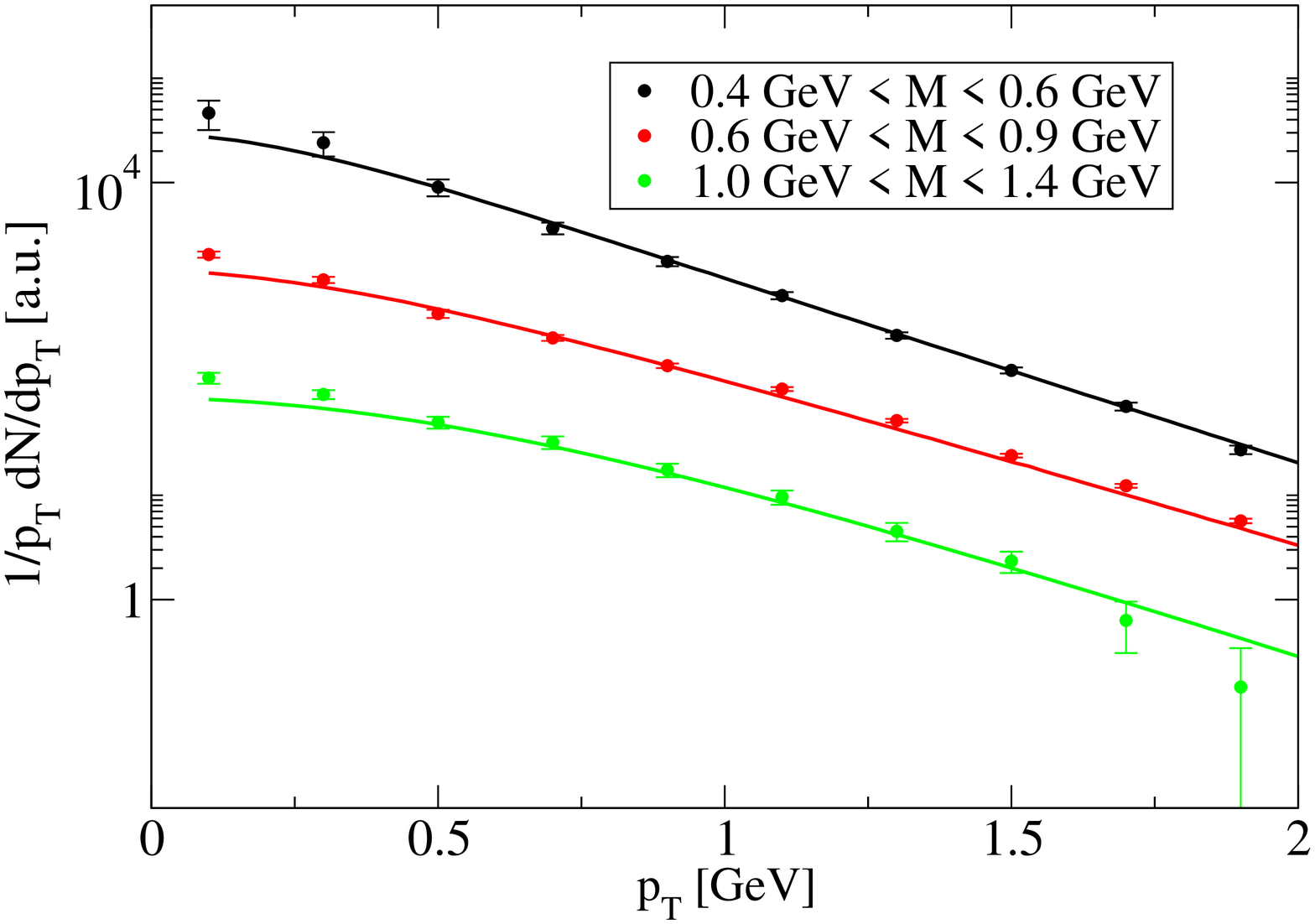, width=6.0cm} \hspace{0.2cm}
\epsfig{file=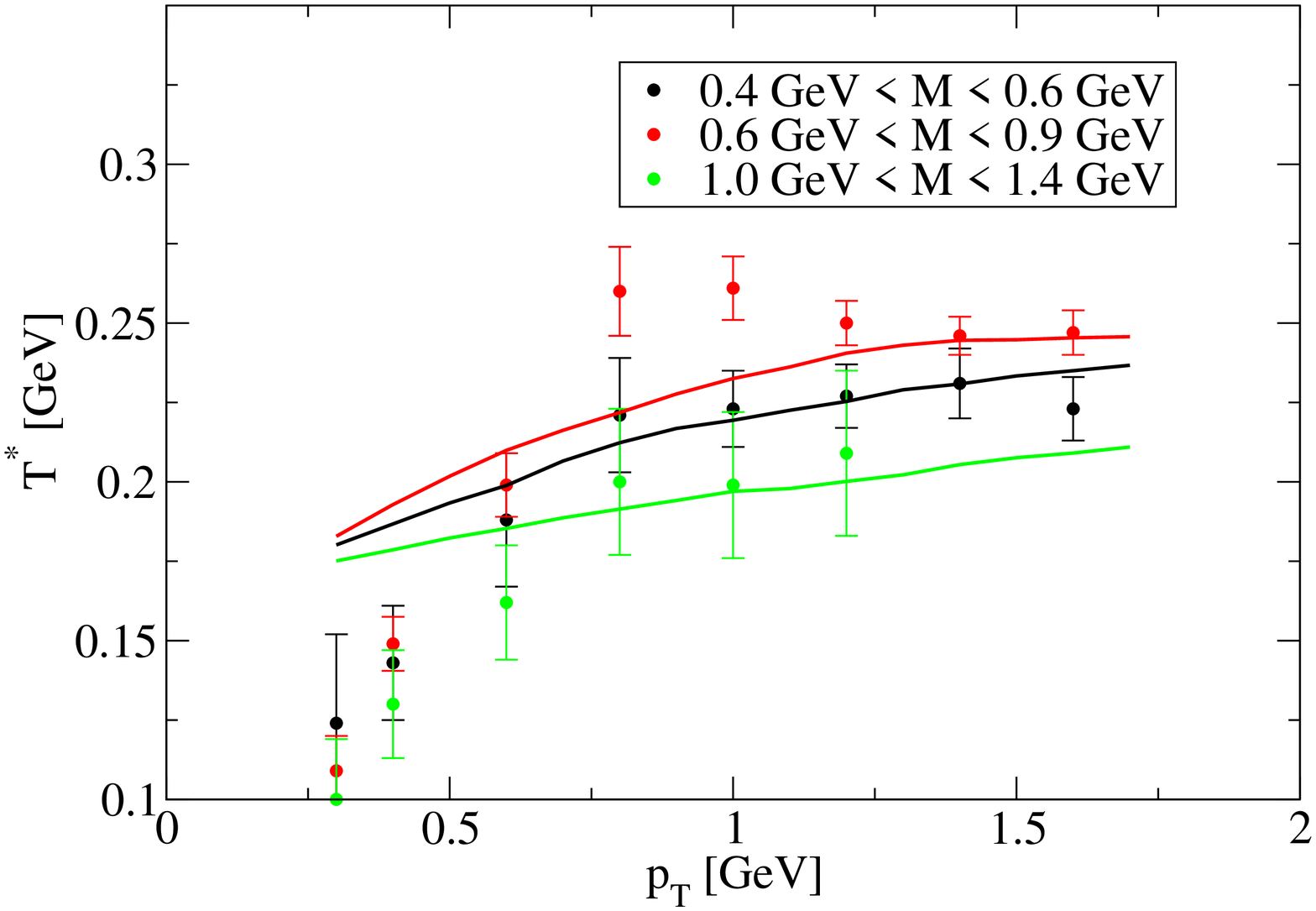, width=6.0cm} 
\end{center}
\caption{\label{F-SPEC} Left panel: $p_T$-spectra in the three different mass windows as obtained by NA60 \cite{NA60data} as compared to the model calculations \cite{ourModel, ourModelPT}. right panel: Effective temperature of the spectra determined from an exponential fit for a $p_T$ window of $ 0.8~ {\rm GeV}$ ($0.6~{\rm GeV}$ for the first data point) as a function of the window center.}
\end{figure}
There is a clear discrepancy between  the model's prediction and the data in the low $p_T<0.5~{\rm GeV}$ region for all masses. Experimentally values of $T^*$ as low as about $100~{\rm MeV}$ are observed. This cannot be accounted for by flow characteristics. Without flow and assuming a thermal emission the source would have a temperature of  about $T<100~{\rm MeV}$, with flow the temperature in the radiating volume would be even lower. The strict anticorrelation of $T$ and flow in any hydrodynamically motivated model makes it very questionable that this low effective $T^*$ contribution could result from any thermal source. We will therefore not address the region below $p_T<0.5~{\rm GeV}$ in the following discussion.
The lower mass and the $\rho$-like region receive dominant contributions from the late hadronic evolution stages, especially from regions  with temperatures $T<170~{\rm MeV}$ and the vacuum $\rho$-contribution.  This implies that the blueshift of the spectra's effective temperature by flow is large - it amounts to about $\approx 70~{\rm MeV}$ which is larger than the maximum temperature variation in the hadronic phase which is only $\approx 40~{\rm MeV}$. In particular it is larger than the changes in effective temperature caused by the $p_T$ dependence of the spectral function in the hadronic phase - even neglecting any $p_T$ dependence of the spectral function would amount to a change of less then $\approx 10~{\rm MeV}$ in slope. The difference between the low mass and the $\rho$-like region is caused by the contribution of the vacuum $\rho$ which receives the maximum amount of flow and predominantly contributes to the $\rho$-like region at $p_T > 1~{\rm GeV}$. We find $T^* \approx 0.27~{\rm GeV}$ for that contribution. This puts severe experimental constraints on the flow field close to freeze-out and by backward extrapolation on the transverse and longitudinal flow fields also at earlier times.
The interested reader can find a result of the calculation of $T^*$ as a function of $M$ in Fig. 2 of \cite{ourModelPT} which illustrate even more clearly the significant differences in $T^*$ as a function of mass. The prominent role of the vacuum $\rho$ contribution is clearly apparent from this figure.
 The higher mass region above $1~{\rm GeV}$ is dominated in our evolution model by contributions from high temperatures. Since this contribution stems predominantly from early evolution stages in which considerable flow has not yet been built up, the predicted effective temperature is lower than in the low mass and $\rho$-like region. This leads to good agreement with the $T^*$ data in the $p_T  > 0.5~{\rm GeV}$ region.
One can turn this argument around and generalize it beyond our specific evolution model: if one assumed that the source were predominantly from cooler hadronic stages, $T  < 170~{\rm MeV}$, and one would allow for adjustments in the flow field in the model to account for the observed $T^*$ in the higher mass region hadronic contributions would have 
to cease rather suddenly at around $\approx 160~{\rm MeV}$, otherwise the strong flow field (required by the lower mass bins which are dominated by emission from those regions) would push the $T^*$ above the data. Even if this fine tuning were performed and hadronic sources in the range of $150~{\rm MeV}<T<160~{\rm MeV}$ contributed dominantly the substantial flow field needed is a strong indication for a higher temperature precursor state.
This especially puts strong constraints on claims that four-pion annihilation processes might be the dominant source  in the higher mass dimuon region \cite{Rapp}. They can have considerable contributions to the yield only from emission at temperatures close to $T_f=130~{\rm MeV} \ll 160~{\rm MeV}$ due to the growing fugacity factor $\exp(4\mu_\pi(T)/T)$ with smaller $T$.
Explicit calculations employing four-pion annihilation rates \cite{Lichard} show that the dimuon yield from these processes is not enough to account for the observed yield in the mass region $M >1~{\rm GeV}$ even if fine-tuning of the source is performed to enforce the compatibility of the measured $T^*$ in the higher mass region.
 We point out that this investigation only implies that the {\it dominant} source in the $M>1.0~{\rm GeV}$ 
mass region of the dimuon spectra are not four-pion annihilation processes, it is still possible  that these processes contribute subdominantly. For a more detailed discussion of this issue, see \cite{ourModelPT}.
One can therefore conclude that the significant lower effective temperature in the $M>1.0~{\rm GeV}$ region is a strong argument for a dominant contribution from stages of the evolution which radiate with  high $T>170~{\rm MeV}$ .

\vspace{-0.25cm}

\ack

We thank P.~Lichard for providing his calculation of $4\pi$ annihilation rates and S. Damjanovic, C. Gale, J. Kapusta, P. Lichard, and H. Specht for discussions. This work was supported by the Academy of Finland, project 206024, and by the Natural Sciences and Engineering Research Council of Canada.

\end{document}